 \documentclass[final,1p,times,twocolumn]{elsarticle}

\usepackage{amssymb}

\journal{Journal of the Mechanical Behavior of Biomedical Materials}

\begin{document}
	
	\begin{frontmatter}
		
		
\title{Deformation and fracture of echinoderm collagen networks}  

\author{Markus Ovaska\textit{$^{a}$}, Zsolt Bertalan\textit{$^{b}$}, Amandine Miksic\textit{$^{a}$},  Michela Sugni\textit{$^{c,d}$}, Cristiano Di Benedetto\textit{$^{c}$}, Cinzia Ferrario\textit{$^{c}$},
	Livio Leggio\textit{$^{c}$}, Luca Guidetti\textit{$^{c,d}$},
	Mikko J. Alava\textit{$^{a}$}, Caterina A. M. La Porta $^{\ast}$\textit{$^{c,d}$}, Stefano
	Zapperi$^{\ast}$\textit{$^{d,e,b,f,a}$} }
. 
\address{$^{a}$ Department of Applied Physics, Aalto University, P.O. Box 11100, FIN-00076 Aalto, Espoo, Finland\\
	$^{b}$ Institute for Scientific Interchange Foundation,  Via Alassio 11/C, 10126 Torino, Italy\\
	$^{c}$ Department of Biosciences, University of Milan, via Celoria 26, 20133 Milano, Italy\\
	$^{d}$ Center for Complexity and Biosystems, 
			University of Milan, via Celoria 16, 20133 Milano, Italy.\\
	$^{e}$ Department of Physics, University of Milan, via Celoria 16, 20133 Milano, Italy. \\
	$^{f}$ CNR - Consiglio Nazionale delle Ricerche,  Istituto per l'Energetica e le Interfasi, Via R. Cozzi 53, 20125 Milano, Italy\\
	$^{\ast}$ Corresponding authors: caterina.laporta@unimi.it, stefano.zapperi@unimi.it}

\begin{abstract}
	Collagen networks provide the main structural component of most tissues and
	represent an important ingredient for bio-mimetic materials for bio-medical applications. Here we
	study the mechanical properties of stiff collagen networks derived from three different echinoderms 
	and show that they exhibit non-linear stiffening followed by brittle fracture. 
	The disordered nature of the network leads to strong sample-to-sample fluctuations in 
	elasticity and fracture strength. We perform numerical simulations of a three dimensional model for the deformation 
	of a cross-linked elastic fibril network which is able to reproduce the macroscopic features of the 
	experimental results and provide insights on the internal mechanics of stiff collagen networks. 
	Our numerical model provides an avenue for the design of collagen membranes with tunable
	mechanical properties.
\end{abstract}


\end{frontmatter}

\section{Introduction}

Living systems provide a formidable source of inspiration for the design and synthesis of new classes of materials 
with a potentially wide range of medical and non-medical applications \cite{huebsch2009,wegst2015}. The structural 
and mechanical properties of animals mostly relies on collagen \cite{buehler2006}, one of the main protein constituents 
of the extracellular matrix in connective tissues and an essential
component of mammalian skin, bones and tendons. To achieve remarkable mechanical properties, long collagen molecules self-assemble into  hierarchical structures by forming fibrils that can then arrange into bundles, known as fibers, and 
extended elastic networks. The mechanics of fiber networks has received a wide attention both experimentally  \cite{vader2009,lindstrom2010,lindstrom2013,munster2013,licup2015} and computationally \cite{Qin2011,das2012,lee2014,feng2015,zagar2015,Depalle2015,feng2016} because of its relevance for applications
but also as a model system to understand soft and disordered matter in general. 

Most of the experimental studies focus on shear deformation of very soft collagen gels 
which are typically extremely soft at small deformations and become stiffer as the deformation proceeds
\cite{motte2013}.  This peculiar non-linear strain stiffening, observed widely in soft collagen gels under shear \cite{vader2009,lindstrom2010,lindstrom2013,munster2013,licup2015} but also present in stiffer
 skin tissues under tension  \cite{yang2015},  has important mechanical and physiological implications since it provides at the same time stability for large deformations and facilitates soft remodeling at small deformations. The precise origin of non-linear stiffening is still under scrutiny. In biopolymer networks this general behavior is usually attributed to a crossover between bending dominated and stretching dominated elasticity \cite{onck2005} or to the strain induced reorientation of the fibers \cite{vader2009,feng2015}. The application of these ideas to collagen is still debated \cite{licup2015} and it is not clear if the deformation is dominated by the non-linear response of individual fibers or by collective properties of the network \cite{gardel2004}.

Marine-derived biomaterials in general --- and marine collagens in particular --- are promising
materials for applications to tissue engineering. The mutable collagenous tissues (MCTs) of echinoderms, widespread in all  five extant echinoderm classes, display striking passive mechanical properties (viscosity, tensile strength, and stiffness) that can be actively transformed within seconds by active mechanisms controlled by the nervous system \cite{wilkie2005}. Hence, in addition  to the mechanical functions usually associated with ‘conventional’ collagenous structures (i.e. energy storage, transmission and dissipation), MCTs allow for the detachment of appendages or body parts in response to disease, trauma or predator attack \cite{wilkie2005} and for the energy-sparing maintenance of posture \cite{takemae2009}. As in mammalian connective tissues, most mutable collagenous structures are formed by  
parallel aggregations of collagen fibrils attached to proteoglycans \cite{trotter1994,sugni2014}. Collagen fibers are 
usually surrounded by an elastic network further improving the structural and mechanical properties
of the tissue \cite{wilkie2005}. 

\begin{figure}[thb]
\centering
  \includegraphics[width=12cm]{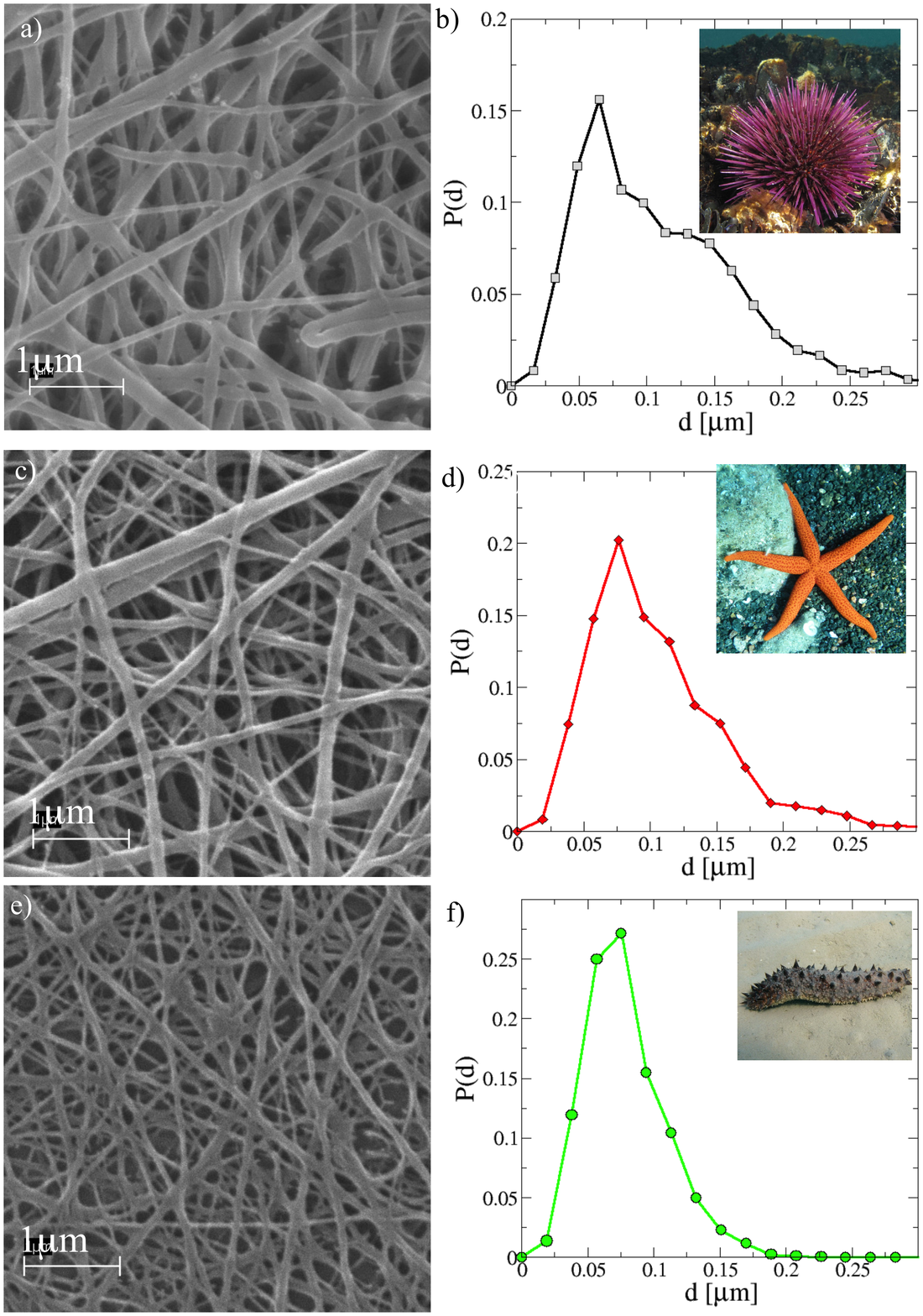}
  \caption{{\bf Echinoderm collagen networks.} The fibril structure of collagen networks as observed by SEM are reported in panels a),c),e) for sea urchin ({\it Paracentrotus lividus}) (inset of b, photo by Federico Betti) starfish ({\it Echinaster sepositus}, (inset of d, photo by Tato Grasso, CC 3.0) and sea cucumber ({\it  Holothuria tubulosa} (inset of f, photo by Roberto Pillon, CC 3.0). The corresponding diameter distributions are reported in panels b),d),f).}
  \label{fig:collagen}
\end{figure}

Besides their physiological relevance, MCTs have been recently proposed as a promising source of collagen for applications to regenerative medicine \cite{dibenedetto2014,sugni2014}. In particular, collagen barrier-membranes of mammalian origin are typically used to facilitate proper tissue regrowth in anatomically separated compartments during guided tissue regeneration \cite{glowacki2008}. Marine collagen and MCTs in particular could provide an 
innovative and safer alternative to mammalian ones \cite{silva2014,dibenedetto2014,ferrario2016}. 
Moreover, other new intriguing applications could be derived from the use of collagen derived from echinoderms: for example to investigate cellular contractility as a biomechanics-related readout for phenotypic drug assessment \cite{zhao2015}

In this paper, we characterize the elastic tensile deformation and failure of  
cross-linked collagen networks derived from three echinoderms (Fig. \ref{fig:collagen}): 
sea urchin ({\it Paracentrotus lividus}), starfish ({\it Echinaster sepositus}) 
and sea cucumber ({\it  Holothuria tubulosa}).  We analyze the stiffening behavior of the 
networks and characterize sample-to-sample fluctuations in the elastic modulus and tensile strength. We compare our results with numerical simulations for a three dimensional model for cross-linked elastic fibrils with parameters chosen to mimic the
microstructure of our samples as revealed by SEM microscopy. Our model is similar in spirit to 
recent three dimensional models of cross-linked fiber networks \cite{Qin2011,Qin2013,lee2014,Depalle2015,zagar2015} but 
its features and parameters are adapted to the stiff collagen networks we study experimentally.
In particular, we explicitly endow each cross-linker with bending, shear and torsional rigidities \cite{zagar2015}.  
Simulations of the model highlight the importance of cross-linking in determining the mechanical properties of these networks, setting the bases for a computational guided design for this kind of bio-inspired materials.

\begin{figure}[thb]
	\centering
	\includegraphics[width=\columnwidth]{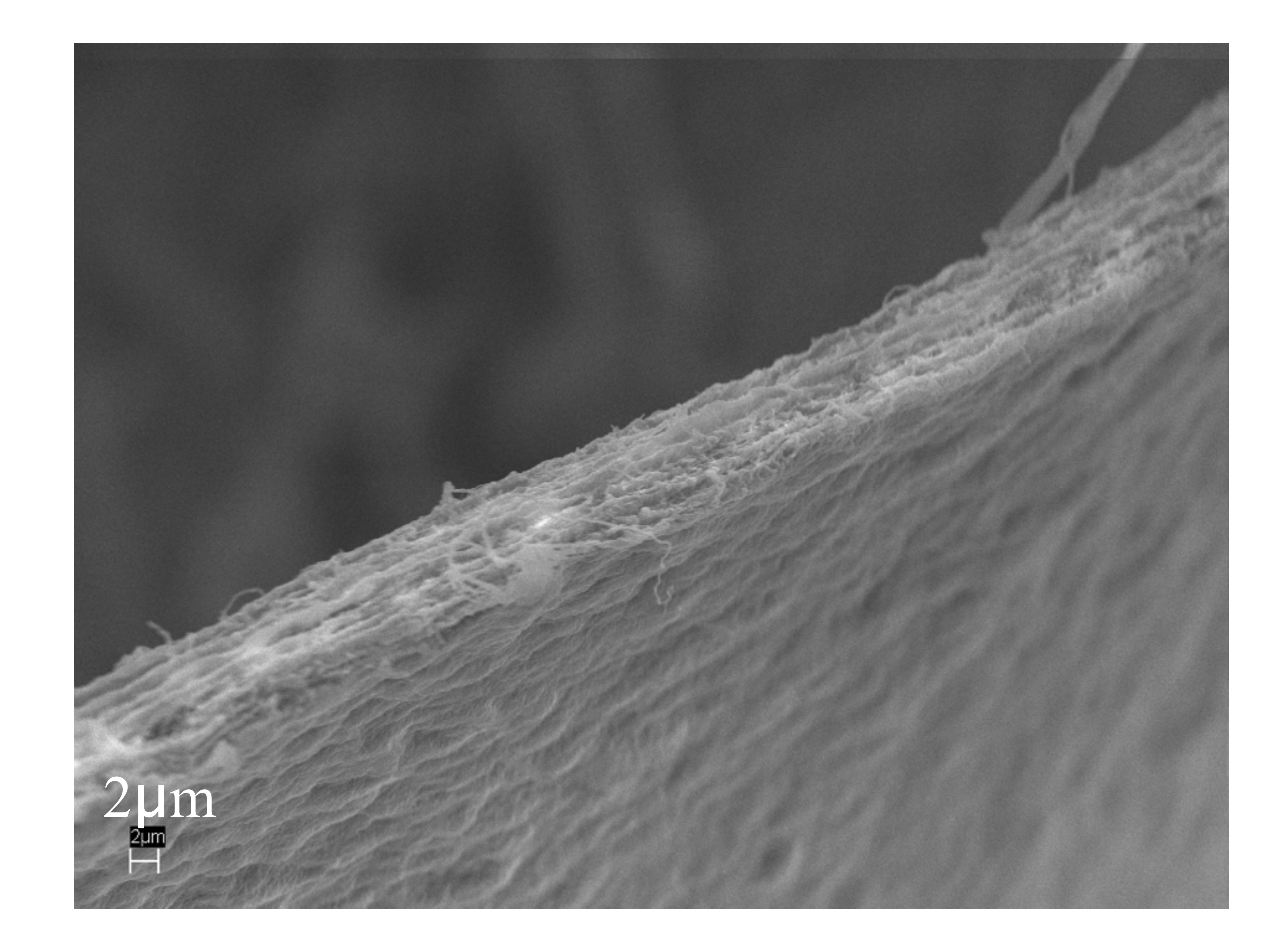}
	\caption{{\bf Fracture surface of a collagen membrane.} A fractured sea urchin collagen membrane as observed by SEM.
		Notice the very smooth fracture surface typical of brittle fracture.}
	\label{fig:fracture}
\end{figure}

\begin{figure*}[thb]
\centering
  \includegraphics[width=12cm]{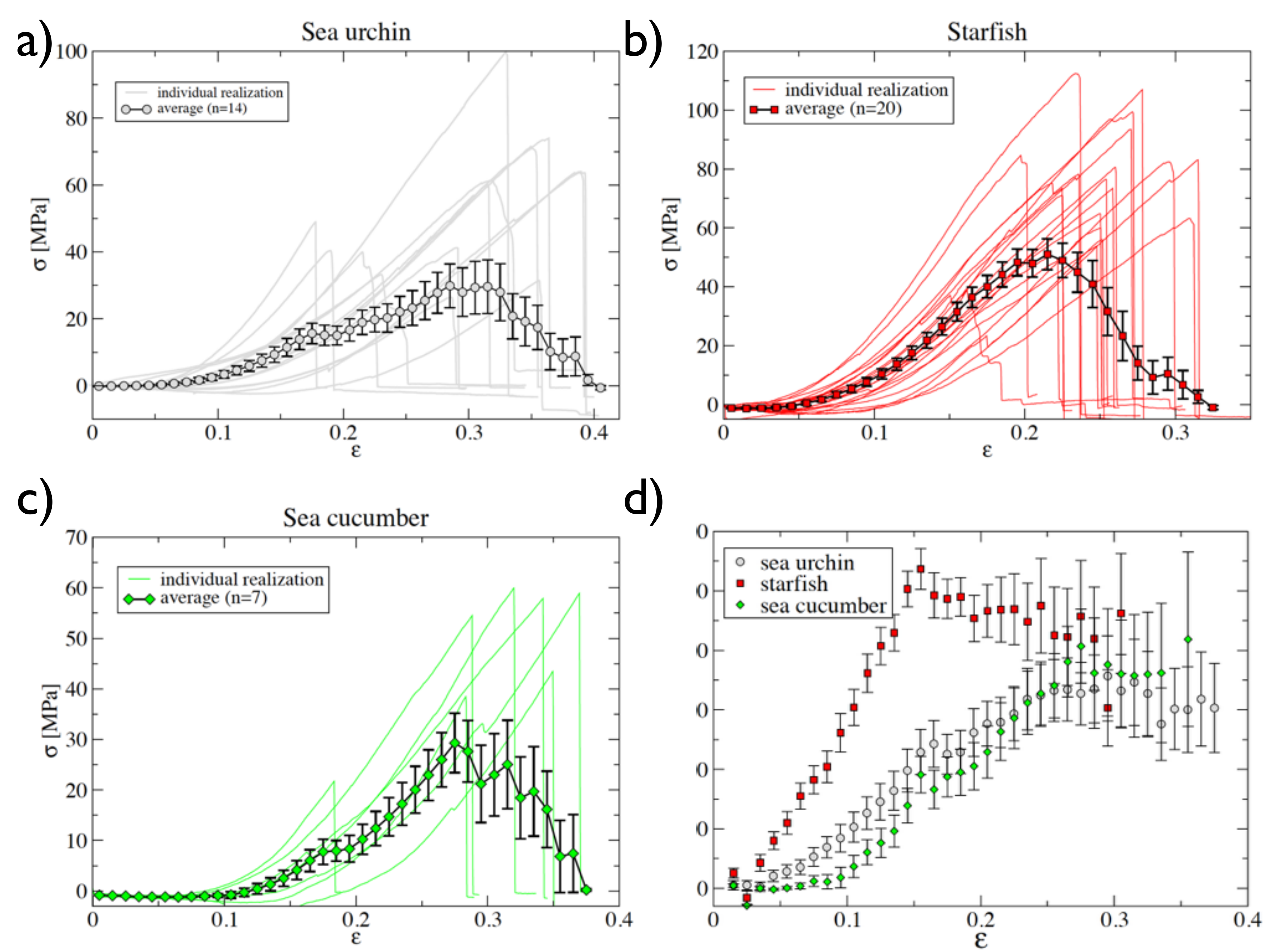}
  \caption{{\bf Elastic response of collagen membranes.} The stress-strain curves measured in collagen membranes for $n$ independent experimental realizations and the corresponding average curves are non-linear and display strain-stiffening. Data reported for a) sea urchin, b) starfish and c) sea cucumber. The Young modulus exhibits an initial strain stiffening and then reaches a plateau value $E^*$ (d). The error bar is the standard error.}
  \label{fig:exp-ss}
\end{figure*}

\section{Experimental methods}

\subsection{Experimental animals}
Adult specimens of the starfish {\it Echinaster sepositus}, the sea urchin {\it Paracentrotus lividus} and the sea cucumber {\it Holothuria tubulosa} were collected by scuba divers in Paraggi (Marine Protected Area of Portofino, Ligurian Sea, Italy) and quickly transferred to the Department of Biosciences (University of Milan) in order to be immediately dissected. The aboral body walls (ABW) of the starfish, the peristomial membrane (PM) of the sea urchins and the whole body walls (BW) of the sea cucumbers were dissected and stored at -20$^\circ$ C for the subsequent collagen extraction protocol.

\subsection{Echinoderm collagen extraction}
Sea urchin and starfish collagen were extracted from the PM and ABW, respectively, as previously described\cite{dibenedetto2014}  with only slight modifications for the latter (see below). Briefly, both starfish ABW and sea urchin PM were dissected in small pieces, rinsed in artificial sea water (ASW), left in a hypotonic buffer (10 mM Tris, 0,1\% EDTA) for 12 h (RT) and then in a  decellularizing solution (10 mM Tris, 0,1\% SDS) for 12 h (RT). After several washings in PBS, samples were placed in a disaggregating solution (0.5 M NaCl, 0,1 M Tris-HCl pH 8,0, 0,1 M $\beta$-mercapto-ethanol, 0,05 M EDTA-Na). Starfish samples underwent an additional washing step in citric acid (1 mM pH 3-4) between the decellularizing and the disaggregating solutions. All these steps were performed under stirring (rotating) conditions. The obtained collagen suspension was filtered and dialyzed against 0,5 M EDTA-Na solution (pH 8,0) for 3 h (RT) and  against dH2O overnight (RT). 
Sea cucumber BW was dissected in small pieces, placed in PBS and gentamicin (40 $\mu$g/mL) and left in stirring condition (RT) for at least 5 days until an homogeneous collagen suspension was obtained, which was then subsequently filtered. 
The suspensions obtained from the three experimental models were stored at -20$^\circ$C until membrane preparation.

\subsection{Echinoderm-derived collagen membrane production}
Membranes of the three different collagen types were prepared as previously described for sea urchin collagen matrices \cite{dibenedetto2014}. The collagen suspensions were centrifuged for 10 minutes at 50x g and then for 20 minutes  at 2000x g. The pellet was re-suspended in autoclaved dH2O (final concentration 2mg/mL) and 8,7mL of this suspension were placed in 5cm x 3 cm rubber silicone molds and left dry overnight at +37$^\circ$ C. The so obtained collagen membranes were then immersed in a EDC/NHS cross-linker solution (EDC 30 mM/NHS 15 mM in MES buffer 100 mM pH 5,5) for 4 h and then washed with PBS, dH2O and ethanol 70\%.

\begin{figure*}[thb]
\centering
  \includegraphics[width=12cm]{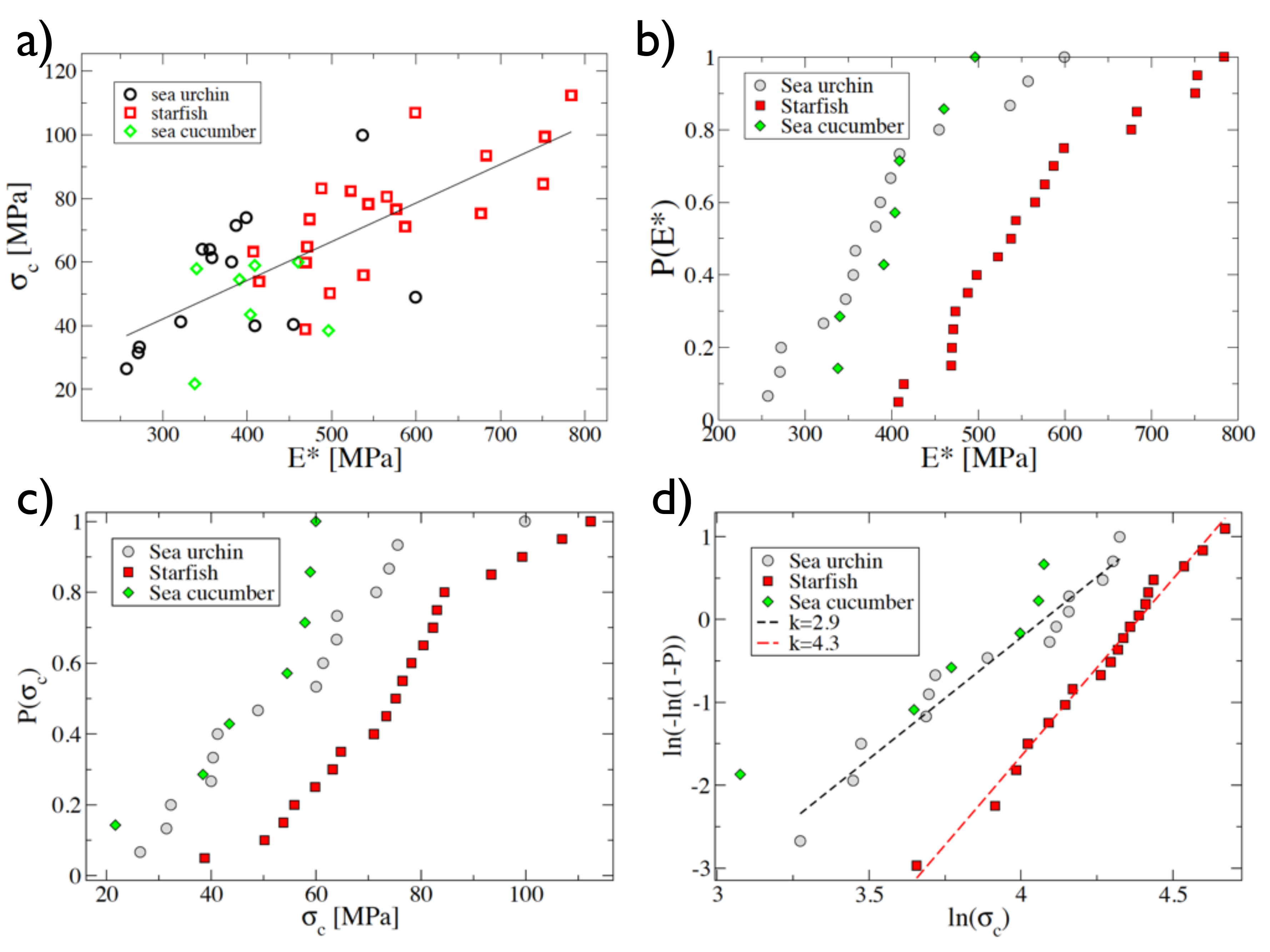}
  \caption{{\bf Stiffness and strength display sample-to-sample fluctuations.} a) We observe considerable 
  sample-to-sample variations in the Young modulus $E^*$ and the tensile strength $\sigma_c$ which appear
  to be correlated (correlation coefficient $r=0.73$, linear least square fit yields a slope $b=0.12 \pm 0.02$ and intercept $a=6\pm 9$ MPa). We also compute the cumulative distributions b) for Young moduli $P(E^*)$ and c)
  tensile strengths $P(\sigma_c)$. d) The strength distribution can be  reasonably fit by the Weibull law.  We observe no significant difference between sea urchin and sea cucumber derived membranes while membranes made out of starfish collagen appear to be stiffer and stronger. }
  \label{fig:exp-stats}
\end{figure*}
 
 \begin{figure*}[thb]
\centering
  \includegraphics[width=12cm]{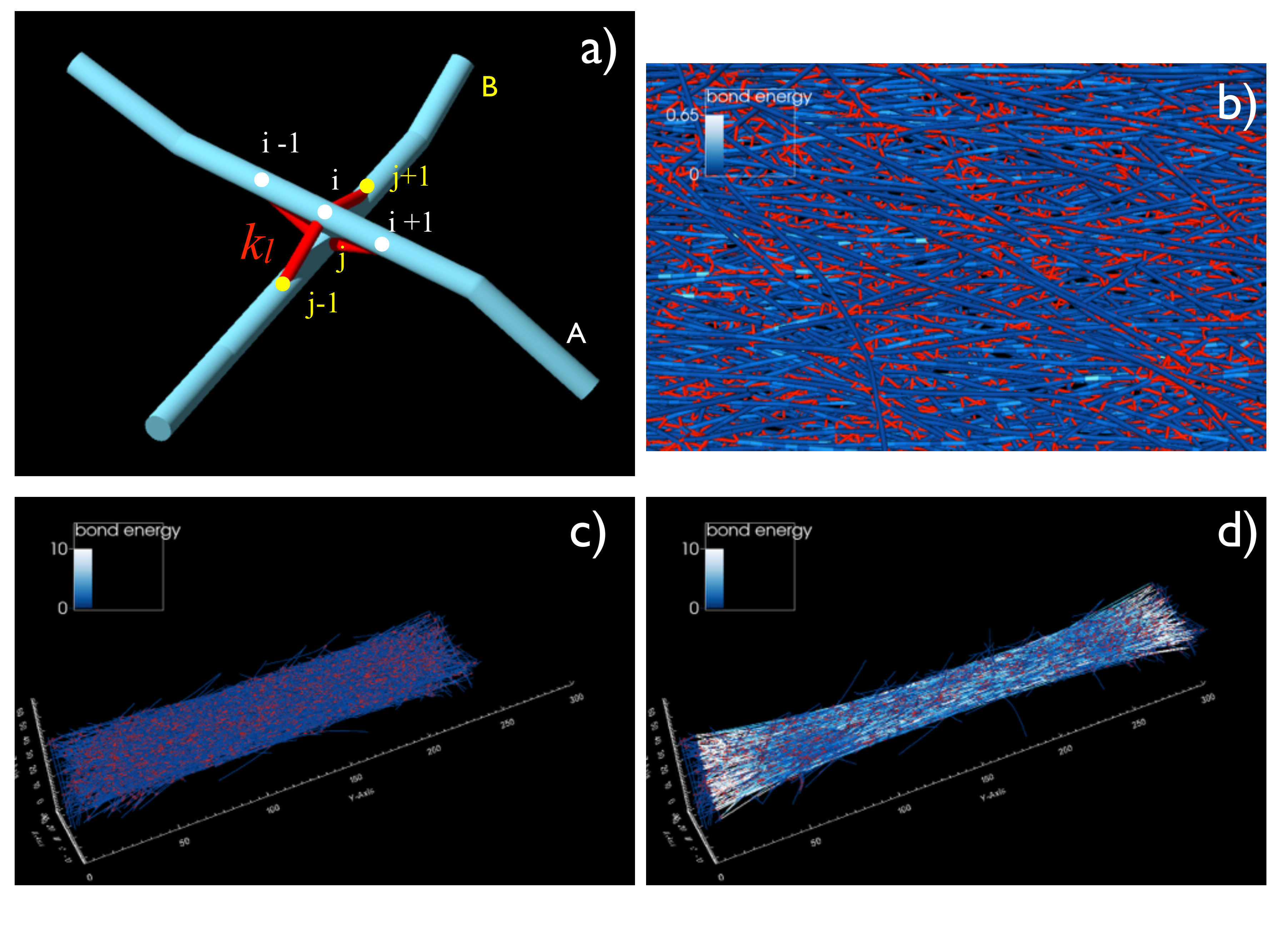}
  \caption{{\bf Cross-linked fibril model.} a) Schematic of the cross-linking between two fibrils A and B. The four cross-linking springs are depicted   in red. b) The initial configuration of the simulated collagen network. A configuration in the (c) non-linear and (d) linear elastic regimes.}
  \label{fig:model}
\end{figure*}

\subsection{Scanning Electron Microscopy}
Substrates were fixed with 2\% glutaraldehyde in 0.1 M cacodylate buffer  for 2 h at 4$^\circ$C and post-fixed with 1\% osmium tetroxide in 0.1 M sodium cacodylate buffer (2 h, room temperature). After several washings with dH$_2$O, they were dehydrated in an increasing ethanol scale and treated with a series of solutions of HMDS (Hexamethyldisilazane) and ethanol in different proportions (1:3, 1:1, 3:1 and 100\% HMDS), mounted on stubs, covered with pure gold (Agar SEM Auto Sputter, Stansted, UK). The samples are observed under a scanning electron microscope (SEM) (LEO-1430, Zeiss, Oberkochen, Germany). Examples of the observed network structure is reported in Fig. \ref{fig:collagen}ace. Images were analyzed using
the ImageJ plugin Diameter-J \cite{hotaling2015} to extract diameter distributions (see  Fig. \ref{fig:collagen}bdf).

\subsection{Tensile tests}
The collagen membranes were cut into narrow strips by pressing them under an assembly
of evenly spaced razor blades. The ends of the collagen strips were then
glued between pieces of rough emery paper to facilitate clamping. The length and width of 
the exposed collagen surface was 10 mm and 2.25 mm, respectively. Before testing, the thickness
of each sample was measured using a LGF-01100L-B transmission-type photo-electric linear encode (Mitutoyo) with 100 nm resolution. Measurements were taken from three positions and averaged. The average thickness was in the range $4.6-18.6 \mu m$. We did not find any systematic variation
of the mechanical properties with the thickness of the sample.
The tensile tests were performed under controlled temperature (21-22 $^\circ$C) and humidity (64-69\%) with a Kammrath-Weiss B.1205. A tensile tester equipped with displacement gauge and DDS3 Controller (Kammrath \& Weiss GmbH, Dortmund, Germany) and a 100 N load cell. After clamping samples into place, Leibovitz L-15 liquid was added on the surface in order to wet them completely. Additional droplets of liquid were added every two minutes during the tests to keep the samples from drying. All samples were tested under uniaxial tension until failure, with a constant strain rate ($8.35 \cdot 10^{-4} s^{-1}$). Fig. \ref{fig:fracture} reports a representative example of a broken sample visualized under SEM microscopy. We observe a very smooth fracture surface, indicative
of brittle failure.


\section{Experimental results}
The individual stress-strain curves obtained experimentally for $n$ equivalent samples are reported in Fig. \ref{fig:exp-ss}a
for sea urchin ($n=14$), in Fig. \ref{fig:exp-ss}b for starfish ($n=20$) and in Fig. \ref{fig:exp-ss}c for sea cucumber ($n=7$).
The curves display a wide variability between sample to sample. The shape of each curve is, however, similar in all cases and
showing a characteristic stiffening behavior with an initial non-linear regime, followed by a linear regime that
ends in abrupt failure, in qualitative agreement with previous experiments on softer mammalian collagen
networks \cite{roeder2002}. To reveal the typical behavior of the system, we compute the average stress strain curve by defining strain bins of size $\Delta\epsilon=0.01$ and computing the average stress in each bin (Fig. \ref{fig:exp-ss}). 
Using the same procedure, we estimate the strain-dependent Young modulus $E(\epsilon)$ as the average slope of the stress-strain curves in each of the strain bins. The result, plotted in Fig. \ref{fig:exp-ss}d, shows a marked linear 
region where $E(\epsilon)$ increases until it reaches a plateau $E^*$. 

To better visualize the large variations observed from sample to sample, we report in Fig. \ref{fig:exp-stats}a the
individual values of the Young moduli $E^*$ and tensile strengths $\sigma_c$, defined as the peak stresses of each curve.
As reported for other natural and synthetic materials \cite{wegst2015},  strength and stiffness are correlated, scaling 
in our case as $\sigma_c \simeq 0.12 E^*$. The sample to sample fluctuations can be quantified by measuring
the cumulative distributions of the Young moduli $P(E^*)$ and of tensile strengths, $P(\sigma_c)$, reported in 
Fig. \ref{fig:exp-stats}b and \ref{fig:exp-stats}c, respectively. The functional form of $P(\sigma_c)$ is well described by the classical Weibull law  $P(x)=1-\exp(-(x/x_0)^k)$ (Fig. \ref{fig:exp-stats}d) as commonly observed in the fracture of disordered media \cite{Alava2006}. Our results indicate that sea urchin and sea cucumber collagen networks display comparable mechanical properties, while starfish derived collagen membranes are slightly stronger and stiffer.

\section{Numerical results}
\subsection{Model}
We construct a three dimensional model of the deformation of cross-linked elastic fibrils.
Collagen fibrils are discretized as bead-spring polymers of diameter $d$ with stretching and bending
stiffness. The rest length of each discretization segment is $a$ and the elastic energy of a fibril of 
$L$ nodes can be written as
\begin{equation}
E_{\textrm{ f}}=\frac{1}{2}\sum_{i=2}^{L}k_{\rm f}(|\textbf{r}_{i}-\textbf{r}_{i-1}|-a)^2+K\sum_{j=1}^{L-1}\cos(\theta_j),\label{eq:f}
\end{equation}
where $k_f=\pi E d^2/4a$ is the stretching stiffness, $K=3 \pi E d^4/64a$ is the bending stiffness, and $\theta_j$ is the angle between two fibril segments. The fibrils are assumed to have no torsional resistance. The Young modulus of an individual fibril is chosen to be $E=10$ GPa, as measured in AFM tests in dry conditions \cite{wenger2007}. 
To simulate fibril-fibril mutual hindrance, fibril nodes also interact by a purely repulsive Lennard-Jones (LJ) potential
$U(r) = 4\epsilon[ (\sigma / r)^{12} - (\sigma / r)^6] $ for $r<\sigma$ and zero otherwise. The parameter 
$\sigma$ is chosen to represent the typical fibril diameter (e.g. $\sigma=75$nm).

To simulate cross-linking, we randomly place links between neighboring fibril nodes. In particular, fibrils are
cross-linked with probability $p$ when two nodes are closer than a distance $b$.
In order to provide the cross-linkers with stretching, bending and torsional stiffness, we model each 
link by a set of four springs (see Fig. \ref{fig:model}a). Considering two fibrils, A and B, a link involves four nodes, two on fibril A, $i-1$ and $i+1$ and two on fibril B, $j-1$ and $j+1$, which are connected by four springs $i \pm 1\rightarrow j\pm 1$ (Fig. \ref{fig:model}a). We notice that internal stresses are present in real samples \cite{Lieleg2011}, but are not easily
quantified in our case. Hence, for simplicity, we assume that initially there is no stress in the system and set the rest length of each of the four link connections to their initial distance $b^\alpha_{i,j}$, with $\alpha=1,...4$. 
The cross-linking energy between fibril $i$ and $j$ is given by
\begin{equation}
E_{\textrm{link}}=\frac{1}{2}k_{\rm l}\sum_{\alpha=1}^4(|\textbf{dr}^\alpha_{\{ij\}}|-b^\alpha_{\{ij\}})^2, \label{eq:link}
\end{equation}
$\textbf{dr}$ is the actual length of the link. In most of the simulations, we assume that both fibrils and cross-linking bonds are linearly elastic and do not consider fracture events or the formation of new cross-linkers. We also perform simulations
where the cross-linkers display non-linear elastic behavior as recently shown in Ref. \cite{Depalle2015}. In particular, the
cross-linker $k_{l}$ stiffness in Eq. \ref{eq:link} is increased from $k_{l,1}=13.3$N/m to a value $k_{l,2}=1330$N/m when the strain in the link is larger than a threshold $\epsilon_1$.

Most simulations are performed with $N=1500$ fibrils randomly arranged in a 7.5 $\times$ 30 $\times$ 3,75 $\mu$m$^3$ block
with open boundary conditions (see Fig. \ref{fig:model}b for a typical starting configuration).
We also perform additional simulations with $N=750$ and $N=12000$ fibrils, with all the dimensions of the simulation box doubled in the latter case. Other geometrical parameters, estimated on the basis of the SEM micrographs, are reported in Table 1.  We deform the sample by holding a set of nodes on one edge of the sample fixed and moving nodes at the other edge at constant velocity. The applied strain-rate for all simulations is 
chosen to be $\dot{\epsilon}_{yy}=10$s$^{-1}$.  As it is common in molecular dynamics simulations, we can only
simulate strain rates that are much larger than the experimental ones. The simulations are performed with a 
linear damping term ($F=-\gamma v$), with damping coefficient $\gamma = 2 \times 10^{-14}$ kg/s and the mass of each element
is set to $M=10^{-18}$ kg, consistent with the estimate of Ref. \cite{buehler2006} for a smaller volume element. The model is  implemented in the LAMMPS molecular dynamics simulator package~\cite{Plimpton1995} and the integration time step is 
set to $dt=6 \times 10^{-8}$s.

\begin{table}
	\centering
		\begin{tabular}{lccc}
			\hline 
			Parameter & Symbol  & SI value\\
			\hline 
			\hline 
			fibril segment length & $a$ & 150 nm\\
			\hline 
			fibril length & $50a$ & 7.5 $\mu$m\\
			\hline 
			fibril diameter & $d$  & 75 nm\\
			\hline 
			Young's modulus & $E$  & 10 GPa\\
			\hline 
			bending stiffness  & $K$  & $3.1\times10^{-13}$ Nm\\
			\hline 
			stretching stiffness & $k_{f}$  & 294.5 N/m\\
			\hline 
			link stiffness & $k_{l}$ & 13-1330 N/m\\
			\hline 
			link distance & $b$  & 150-450 nm\\
			\hline 
			link probability & $p$  & 0.1-1.0\\
			\hline 
			volume fraction & $\rho$  & 0.03-0.06\\
			\hline 
			number of fibrils & $N$  & 750-12000\\
			\hline 
			repulsion energy & $\epsilon$ & 1.65$\times 10^{-13}$ J \\
			\hline 
			repulsion radius & $\sigma$ & 75 nm \\
			\hline
			damping parameter & $\gamma$ & $2\times 10^{-14}$ kg/s \\ 
			\hline 
			fibril segment mass & $M$ &  $10^{-18}$ kg
		\end{tabular}
	\caption{Parameters used in the numerical simulations in  SI units. Notice that the parameter ranges
		(e.g. in $k_l$) represent the range that has been explored in different simulations. \label{tab:sim-params}}
\end{table}

\subsection{Simulations}

\begin{figure}[h]
	\centering
	\includegraphics[width=\columnwidth]{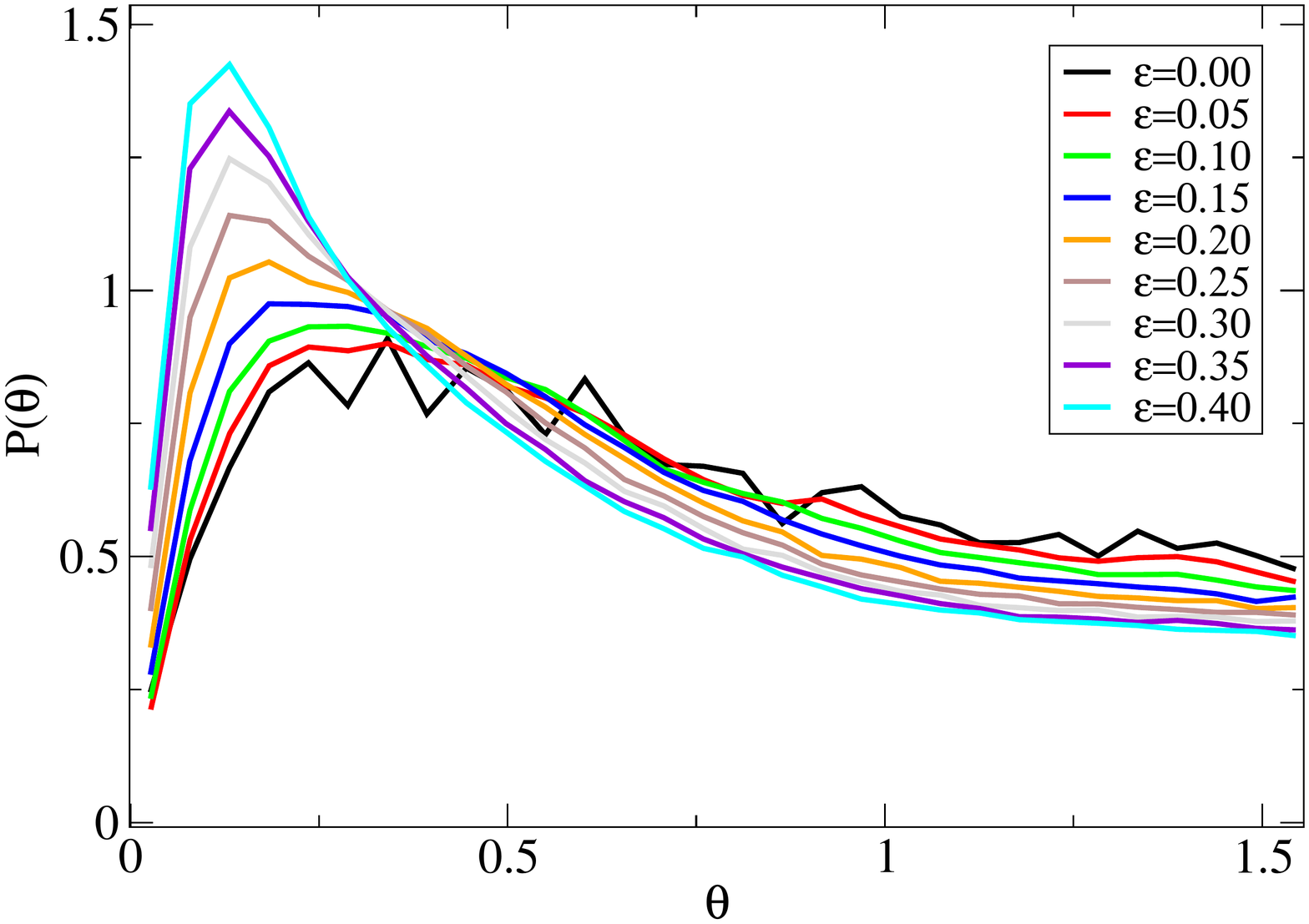}
	\caption{{\bf Fibrils reorient during deformation.} The distribution of fibril angles $\theta$ as a function
	of strain for $N=12000$ fibrils. The data indicate that the angles between between the loading axis and the fibrils 
	decrease with strain.		\label{fig:orient}}
\end{figure}

We perform numerical simulations of the network model using the parameters reported in Table 1. The evolution
of the deformation process is reported in Fig. \ref{fig:model}c-d and in video S1. We observe that fibrils are first 
reoriented in the initial stage and then the sample becomes much thinner. This is quantified in Fig. \ref{fig:orient}
displaying the evolution with strain of the distribution of the angle $\theta$ formed by the tensile axis and fibril segments. Fig. \ref{fig:orient} shows that as the sample deforms the typical value of $\theta$ decreases, as shown by
the left-ward shift of the peak of the distributions. We checked that the effect is more marked when the sample size is larger. 

\begin{figure*}[thb]
	\centering
	\includegraphics[width=12cm]{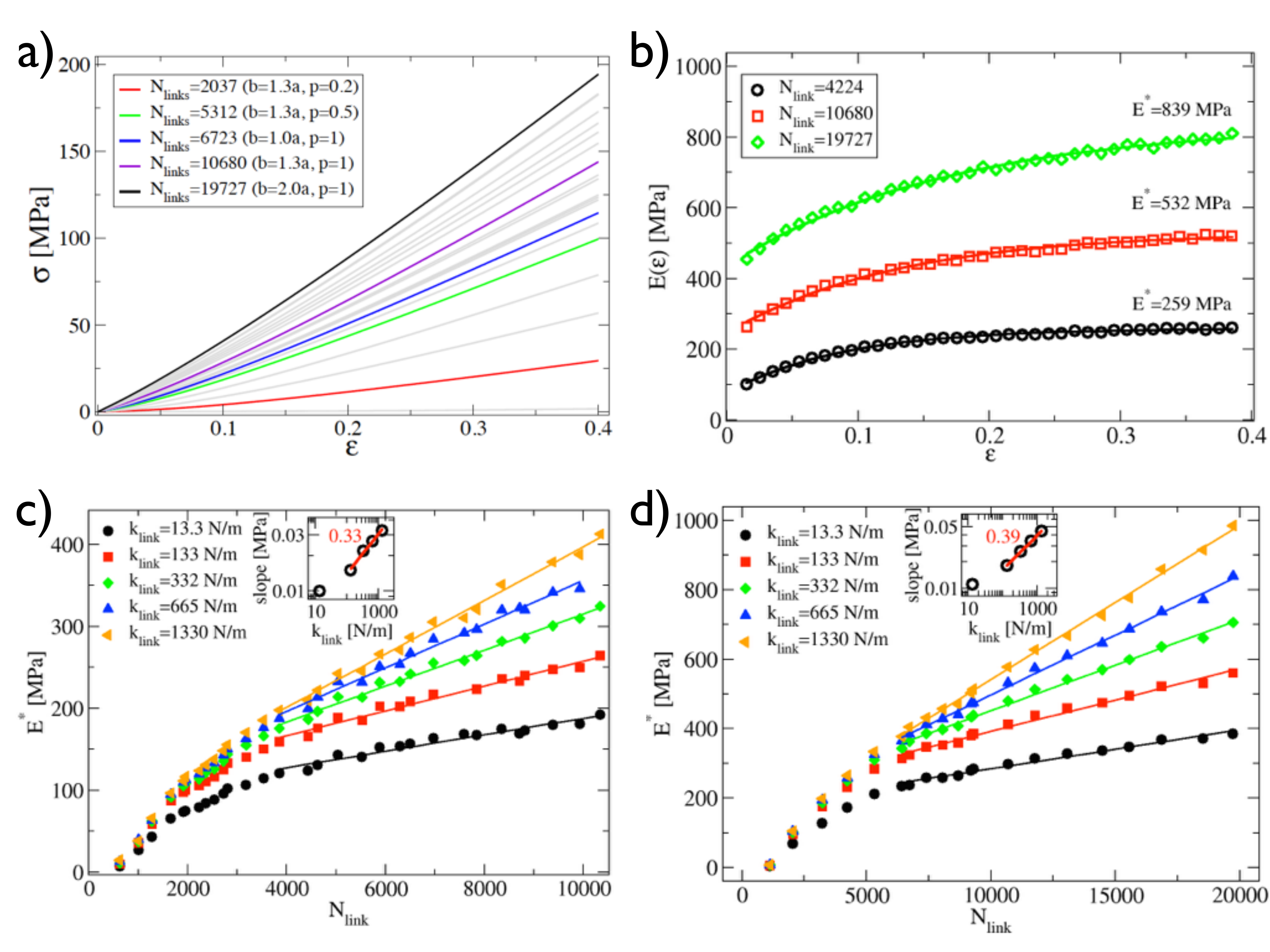}
	\caption{{\bf Mechanical role of the cross-linking probability.} a) Stress-strain curves for different
		values of the number of cross-linkers. b) The strain dependent Young modulus is fit with exponential
		functions (see text). c) The steady-state Young modulus $E^*$ as a function of $N_{link}$ for c) $N=750$ and
		d) $N=1500$ fibrils. The inset shows the dependence of slope with $k_{link}$.
		\label{fig:p-depend}}
\end{figure*}

\begin{figure*}[thb]
	\centering
	\includegraphics[width=12cm]{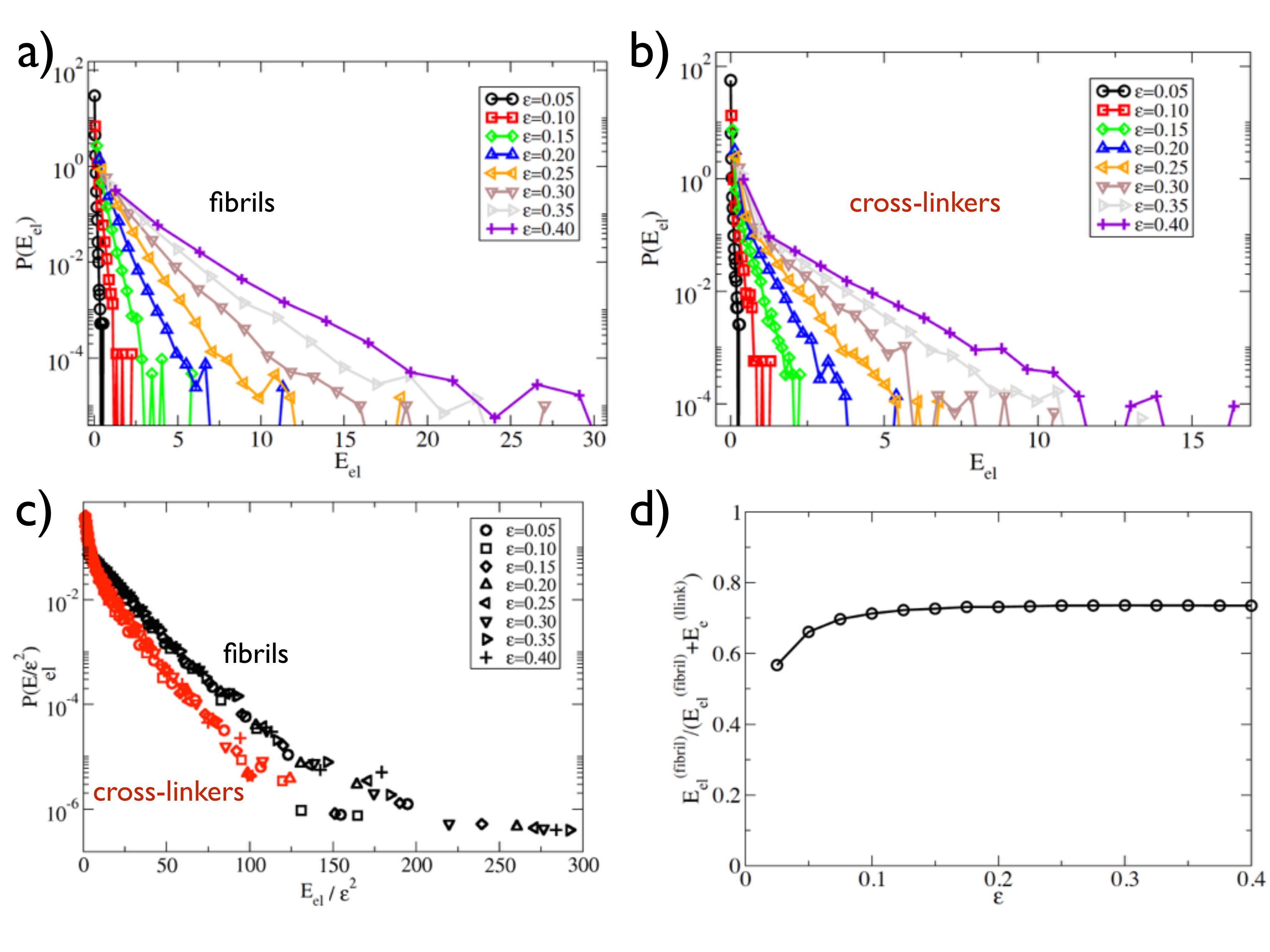}
	\caption{{\bf Elastic energy distributions.} The distribution of elastic energies for a) fibrils and b) cross-linkers
	as a function of strain. c) The distributions in a) and b) can be simply rescaled in terms of $\epsilon^2$. d) The fraction of elastic energy carried by fibrils increases as a function of strain. 
		\label{fig:energy}}
\end{figure*}

From the simulations, we notice that the amount of cross-linking plays a major role in the mechanical response. 
This is illustrated in Fig. \ref{fig:p-depend}a  where we show that the stress-strain curves change
substantially by varying the number of cross-linkers $N_{link}$. The same curves display features that are in 
qualitative agreement with experiments (Fig. \ref{fig:exp-ss}). For each curve we compute 
the local slope that we fit with an exponential function of the strain $E(\epsilon)= E_0+(E^*-E_0)(1-\exp(-\epsilon/\epsilon_0))$, which allows to estimate the steady state Young modulus $E^*$ (Fig. \ref{fig:exp-ss}b). Fig. \ref{fig:p-depend}cd shows that the resulting Young modulus $E^*$ increases with the number of cross-linkers $N_{link}$ and also depends on their stiffness $k_{link}$. The results reported in Fig. \ref{fig:p-depend}c correspond to a set of $750$ fibrils, while those in Fig. \ref{fig:p-depend}d correspond to $1500$ fibrils. Finally, we investigate how the elastic energy is distributed
in the system as it is loaded. In Fig. \ref{fig:energy}ab, we report the distribution of elastic energies $E_{el}$ of fibrils (Eq. \ref{eq:f}) and cross-linkers (Eq. \ref{eq:link}).  Both distributions decay exponentially with a decay constant that simply scales with the loading strain energy (e.g. $E_{el}\propto\epsilon^2$) as illustrated in Fig. \ref{fig:energy}c.
Finally Fig. \ref{fig:energy}d shows the fraction of elastic energy carried by fibrils. As the sample is loaded
a part of the load carried by the cross-linkers is transferred to the fibrils, in agreement with the observation
of a reorientation of the latter. 

\begin{figure*}[thb]
	\centering
	\includegraphics[width=12cm]{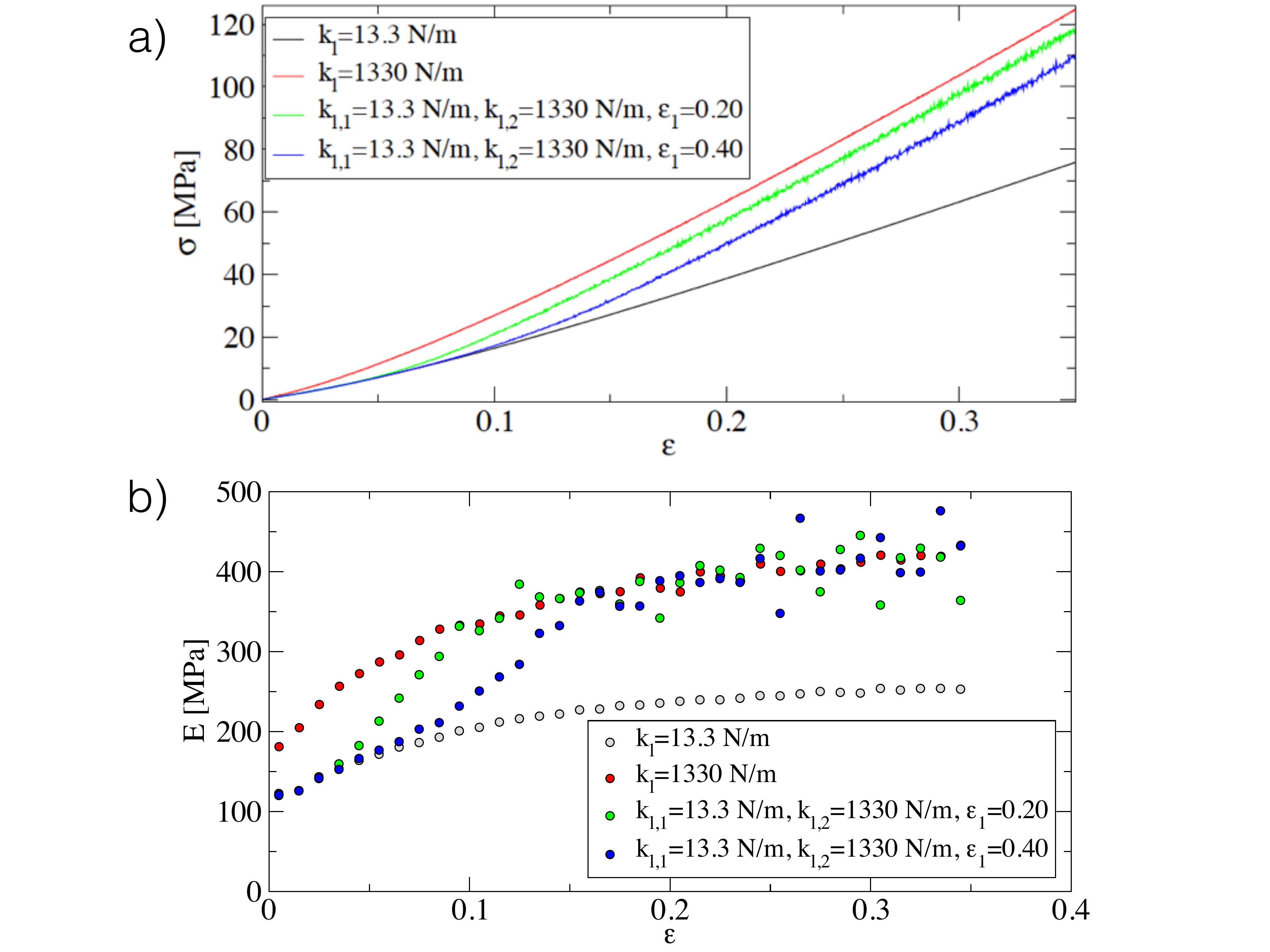}
	\caption{{\bf Role of the non-linear elasticity of cross-linkers.} a) Simulated stress-strain curves for a model
		with non-linear cross-linker springs are compared with the ones obtained with linear cross-linker springs. b) 
		The corresponding strain dependent Young moduli. 
		\label{fig:non-linear}}
\end{figure*}

We notice that the non-linearity of the simulated stress-strain curves and of the resulting Young moduli 
is less pronounced than in experiments (compare  Fig. \ref{fig:exp-ss}b with Fig. \ref{fig:p-depend}b). 
This suggests that geometrical effects due to fibril arrangement and reorientation are not enough to 
fully account for the experimentally observed stiffening. To overcome this problem, we explicitly include
non-linear effects in the cross-linker elasticity following Ref. \cite{Depalle2015}. Fig. \ref{fig:non-linear}a
shows the stress-strain curves obtained for two values of the linear cross-linker spring stiffnesses (i.e. $k_l=13.3$ N/m and $k_l=1330$N/m). The curves are then compared with simulations with non-linear cross-linker springs which interpolate
between the two linear spring stiffness values. The resulting stress-strain curves displays slopes 
that interpolates between the curves obtained with linear cross-linker springs. The local Young moduli, shown
in Fig. \ref{fig:non-linear}b, now display a more pronounced non-linearity that is closer to experimental 
results.

\section{Conclusions}
In this paper, we investigate experimentally and computationally the mechanical properties of
collagen networks. We focus on echinoderm-derived collagen networks, an alternative to
the most widely used collagen of mammalian origin, and show that their mechanical properties 
are very similar to the latter. We notice that for the three types of collagen networks the resulting 
mechanical properties are similar, with Young moduli in the 100-500 MPa
range and tensile strength in the 20-50MPa range, values that are remarkably close
to what is measured in the human skin \cite{gilchrist2012}. This feature is possibly relevant
in order to use marine collagen in biomedical applications \cite{dibenedetto2014,sugni2014}.
Slight quantitative differences exist for starfish collagen that appears to be stronger and stiffer
than the other two. This could be attributed to different cross-linking properties due
to differences in the sample preparation and extraction.
Furthermore, as in other fiber networks studied in the past \cite{vader2009,lindstrom2010,lindstrom2013,munster2013,licup2015},
our samples also display non-linear stiffening and 
sample-to-sample strength fluctuations typical of systems with a heterogeneous microstructure
\cite{Alava2006}. 

To clarify the microscopic origin of the experimentally observed behavior,
we develop a computational model for the deformation of cross-linked elastic fibril networks. 
The model allows to reproduce some key features of the experimental results such as the 
non-linear stiffening behavior. If we use linear-elastic cross-linkers, we find non-linear
stress-strain curves induced by a reorientation of the fibrils. The resulting non-linear
strengthening is, however, weaker than the one measured experimentally suggesting that
other mechanisms playing a role. When we include non-linear elastic cross-linkers in the
model, as suggested by Ref. \cite{Depalle2015}, we obtain a more pronounced stiffening.
Additional sources of non-linearity may come from non-linear elastic deformation of 
the fibril themselves \cite{buehler2006} and from internal stresses due to sample preparation
\cite{Lieleg2011}. 
Furthermore, our model highlights the crucial
role played by the amount of cross-linking in determining the mechanical response of the sample.
This suggests an interpretation from the experimentally measured sample-to-sample fluctuations in the stress strain curves displaying large variations (around 50\%) in the Young modulus. We have performed simulations of the model using the same parameters but different initial arrangements for the fibrils. When we do this we find much smaller variations of the resulting Young modulus (around 5\%). A possible explanation is that   the experimental variations reflect internal fluctuations in the cross-linking probability or in their stiffness, quantities that we have held constant in the simulations but have shown (Fig. \ref{fig:p-depend}) to have a crucial influence on the resulting Young modulus.

Our modeling strategy allows us to obtain an agreement with experimental results and suggests
mechanisms for the observed non-linear elasticity. The current version of the computational
model is presently not able to simulate fracture, since simply breaking overstretched cross-linkers or fibrils leads to discontinuities that are difficult to deal with in the present framework. A possible
solution would involve quasi-static approaches where the network deformation is obtained by energy minimization
as for other models for fracture \cite{Alava2006}. These types of models pave the way for a computationally based design of bio-inspired materials where the mechanical properties can first be explored {\it in silico} to guide the production
of materials with desired mechanical properties.

\section*{Acknowledgements}
MO, AM and MJA  are supported by the Academy of Finland through its Centres of Excellence Programme (2012-2017) under project no. 251748. ZB and SZ are supported by ERC Advanced Grant n 291002 SIZEFFECTS.  SZ thanks the Academy of Finland FiDiPro progam, project 13282993. CAMLP thanks the visiting professor program of Aalto University where part of this work was completed. The authors are grateful to Lahja Martikainen and Matti Toivonen of the Molecular Materials research group at Aalto University for assistance with the mechanical testing.   

\section*{Authors Contributions}
LG,LL,MS prepared samples. CF,MS performed SEM experiments. AM,MO performed tensile tests. MO analyzed data. ZB,CDB,SZ designed the model. ZB wrote the code. MO performed numerical simulations. CAMLP,SZ wrote the manuscript. CAMLP,MJA,SZ designed and coordinated the project.


\section*{Supplementary information}
{\it Video S1}: a representative example of a simulated fibril network under tensile elongation.

\end{document}